# Revealing atomistic mechanisms of gold-catalyzed germanium growth using molecular dynamics simulations

Zhiyi Wang[1], Jixin Wu[2], Yanming Wang[1, *]

*[1] University of Michigan - Shanghai Jiao Tong University Joint Institute, Shanghai Jiao Tong University, 800 Dongchuan Rd., Minhang District, Shanghai, 200240, P. R. China*
*[2] School of Materials Science and Engineering, Huazhong University of Science and Technology, 1037 Luoyu Rd., Hongshan District, Wuhan, Hubei, 430074, P. R. China.*

*Corresponding author, yanming.wang@sjtu.edu.cn (Y. Wang)

**Abstract**

The vapor-liquid-solid (VLS) method is considered a plausible technique for synthesizing germanium (Ge) nanostructures (e.g. nanowires), which have a broad range of applications due to their unique electronic properties and intrinsic compatibility with silicon. However, crystallization failures and material defects are still frequently observed in VLS processes, with insufficient understanding of their underlying mechanisms due to instrumental limitations for high-resolution in-situ characterizations. Employing an accurate interatomic potential well fitted to the gold-germanium (Au-Ge) phase diagram, we performed molecular dynamics simulations for a systematic investigation on the Au-catalyzed growth process of Ge crystals. From the simulations, relationships were established between the overall Ge growth rate and several main synthesis conditions, including substrate crystallographic orientation, temperature and Ge supersaturation in liquid. The dynamical behaviors of Ge atoms near the liquid-solid growing interface were captured, from which the atom surface stability and exchange rate were estimated for quantifying the atomistic details of the growth. These interface properties were further linked to the surface morphologies, to explain the observed orientation-dependent growing modes. This study sheds new lights into the understanding of the VLS growth mechanisms of Ge crystals, and provides scientific guidelines for designing innovative synthesis methods for similar nanomaterials.

*Keywords:* vapor-liquid-solid growth; molecular dynamics simulation; germanium crystals; liquid-solid interface

I.  **Introduction**

In recent years, low-dimensional semiconductor materials have received substantial attention for applications in various areas, due to their unique electronic, photonic, thermal, electrochemical and mechanical properties[1]. For instance, in field effect transistors (FETs), semiconductor nanowires (NWs), as one classical representative of nanomaterials, enable a better integration of electronic components with an improved sensitivity of the device[2]. In the field of energy conversion and storage, the nanowire structures can incomparably increase the energy density, power density, and cycling performance because of their larger surface-to-volume ratio, extra active sites, and better permeability[3]. Among the candidate materials for semiconductor nanowires, germanium (Ge) has its own advantages[4,5], as it has a similar diamond structure with silicon[6], a higher electron and hole mobilities as well as a larger Bohr exciton radius[7].

Many nanowire synthesis methods have been proposed and intensively studied[8,9,10,11], in order to obtain desired nanowire structures at high precision and efficiency. In particular, the VLS method, as one of the bottom-up synthesis routes, has realized a variety of new materials and morphologies including the Ge NWs, with good control over material crystallinity and chemical properties[12,13,14]. A typical VLS process for catalyzed semiconductor growth can be described as follows: a heated metal catalyst and the gaseous precursor first form a liquid alloy, which has a high adhesion capacity to facilitate further material absorption from the gas phase. When the supersaturated state of the liquid reaches, the target material is extracted and then deposits on the liquid-solid interface, leading to a bottom-up continuous growth of the wire.

The VLS method has been widely used for growing Ge NWs; however, challenges and problems such as nucleation in unwanted orientations, nucleation failures and growth anomalies[15], still remain to be fully solved towards improving the wire quality and yield. This has attracted tremendous research explorations in recent years, for developing thermodynamic and kinetic theories as well as capturing the dynamical details of the growth process. For example, it has been found that the equilibrium composition of the Au-Ge liquid alloy at the tip of Ge nanowires is significantly deviated from that for the bulk system, and could be very sensitive to temperature. It has been shown that a reduced equilibrium concentration and/or an increased supersaturation of germanium in liquid may lead to a reduced critical diameter of germanium nanowires. It has also been demonstrated that various structures ranging from pyramidal nano-islands to uniform core-shell structures can be formed by controlling the concentration of Au[16].

While the above studies have identified the importance of several macroscopic synthesis factors including liquid composition, temperature and surface properties; it is more informative to achieve direct microscopic observations of nanocrystal growth process, which actually have become available with the advancement of characterization techniques and tools such as the optical reflectance spectroscopy[17] and environmental TEM[18]. These in situ characterization methods (especially the environmental TEM) have provided valuable new morphological and crystallographic information for understanding nanomaterials growth. However, it has been shown that electron beams may affect the nucleation and growth processes of solid nanoparticles[19], and may cause damage to the liquid chemical environment[20]. Besides, most of the existing in situ characterization techniques are still operated under restrictive conditions that may not be suitable for observing a VLS growth. Thus, a complete understanding of the VLS mechanisms at the atomic level has yet to be achieved, especially how the macroscopic processing factors affect the atomic behaviors during the growth.

This motivates us to adopt the molecular dynamics (MD) method to study the VLS mechanisms of the Au-Ge system, considering that MD simulations have been used for similar VLS systems such as Au catalyzed silicon growth[21,22,23,24]. It should be noted that one general difficulty in the classical MD approach is the availability of an accurate interatomic potential for the target system[25]. In this regard, we have developed an MEAM potential for the Au-Ge system that was well fitted to the binary phase diagram[26], as an enabler of the systematic investigations in this study. Here we performed MD simulations of Au-catalyzed Ge crystal growth on four common Ge substrate orientations in a range of temperature ($T$) and Ge supersaturation ($\Delta x_{Ge}$) conditions. For each substrate orientation, the Ge growth rate was predicted as a function of $T$ and $\Delta x_{Ge}$ using the simulation trajectories, from which the activation energy was estimated. From statistical analysis, the substrate orientation-dependent activity at the liquid-solid interface was quantified by the stable time of surface atoms. Finally, the liquid-solid interface morphology was evaluated at the atomic scale, to clarify the correlations between Ge growth modes and substrate orientations.

## II. Results

### A. Growth Rates of Ge Crystals

According to the liquidus of the Au-Ge phase diagram predicted by the adopted

MEAM potential[26], we firstly constructed 16 configurations: four common crystallographic orientations of Ge solid substrates ({100}, {110}, {111} and {112}[27]), combined with four Ge equilibrium concentrations, corresponding to temperatures of 800K, 850K, 900K, and 950K respectively. Then by substituting different amounts of Au atoms in liquid, three additional Ge supersaturation conditions were created. Under each given condition, we repeated the simulation three times to improve the statistics, yielding a total number of 192 simulations. Collected from all these simulations, the Ge crystal growth (in the unit of 'nm', see Fig. S1 for more details) versus the simulation time (in the unit of 'ns') is shown as a figure matrix in Fig. 1. Each element of the matrix contains four sets of curves, with their colors representing the initial Ge supersaturations (ΔGe%). For each color, the three lighter curves correspond to the original growth data points, while their average value is emphasized by a darker color. The areas fenced by the curves with the same color are dyed to indicate the variation of the simulation predictions. The top axis ticks {100}, {110}, {111} and {112} specify the crystallographic orientations of their corresponding rows, and the left axis ticks 800K, 850K, 900K and 950K clarify the temperatures of their corresponding columns.

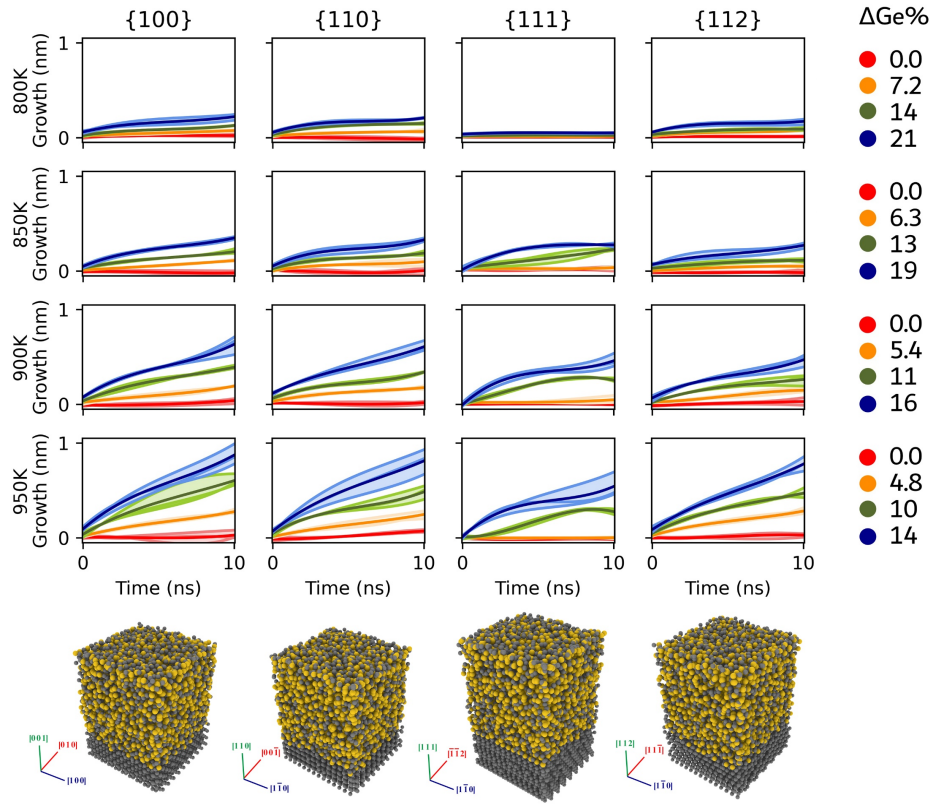

**Figure 1. Height of Ge crystal plotted as a function of time, for simulations performed at four different temperatures, with the solid-liquid atomic configurations of four substrate orientations demonstrated below. The yellow color is assigned to gold atoms, while the germanium atoms are in grey. For each curve, the corresponding initial supersaturation of Ge atoms in the liquid phase (in %) is given on the right.**

From the first derivative of these curves, we obtained a scatter plot of the growth rates for all four crystallographic orientations at 1 ns, 4 ns, 7 ns and 10 ns, which may be described by a general function form inspired by Brice[27] (Eq. 1), where $r$ is the growth rate:

$$r = f(\Delta \text{Ge}) \cdot \exp(-\frac{E_a}{k_B T}) \quad \quad \text{(Equation 1)}$$

The exponential term $\exp(-\frac{E_a}{k_B T})$ describes an Arrhenius-like temperature dependency governed by the activation energy $E_a$ of the Ge crystal growth. The $f(\Delta \text{Ge})$ term is assumed to be a function purely depending on the supersaturation of Ge in liquid. Since both $E_a$ and $f(\Delta \text{Ge})$ should be affected by the crystal orientation of the substrate, further discussions on these factors will be given in Sections II.A.1 and II.A.2 respectively.

1. **Temperature-dependency analysis**

In principle, for a given substrate orientation, when $\Delta \text{Ge}$ is fixed, the growth rate in Eq. 1 can be simplified as $r = C \cdot \exp(-\frac{E_a}{k_B T})$, where $C$ is a constant allowing for a straightforward fitting to obtain the activation energy $E_a$. However, in practice, it is difficult to strictly keeping $\Delta \text{Ge}$ a constant in MD simulations, as $\Delta \text{Ge}$ tends to continuously fluctuate, and changing the number of Ge atoms during the simulation may lead to artificial perturbations to the system. Alternatively, the simulation data points were selected at $\Delta \text{Ge} \approx 14.5\%$, for the curves presented in Fig. 2, where the growth activation energy $E_a$ for each substrate orientation was estimated as summarized in Tab. 1.

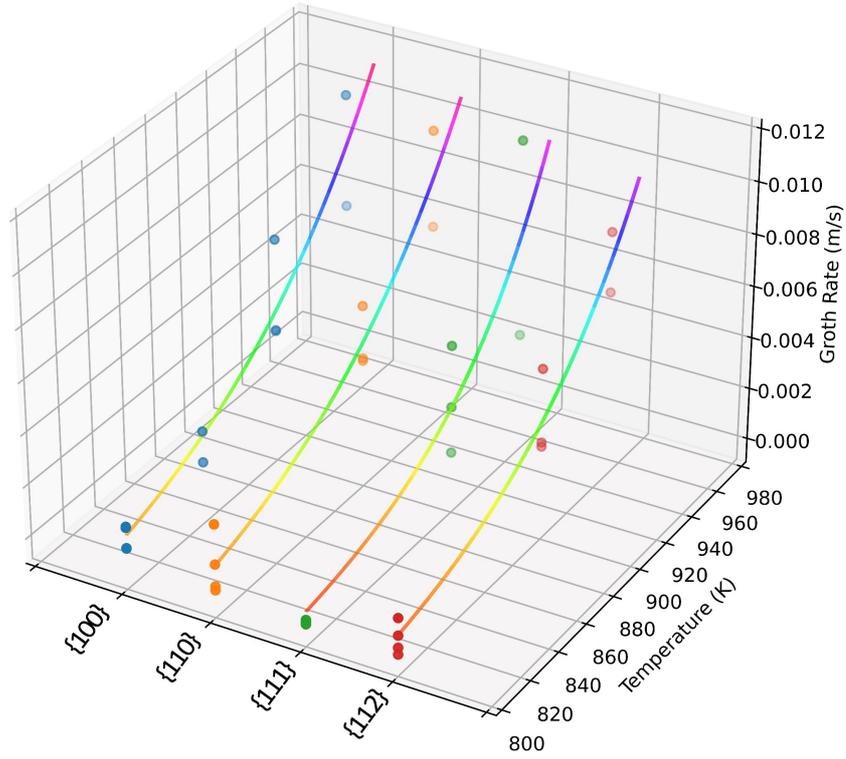

**Figure 2. Fitted temperature-growth rate curves and the original data points for four substrate orientations at ΔGe ≈ 14.5%.**

**Table 1. The activation energy of crystal growth concerning four substrate orientations.**

| Substrate orientation | {100} | {110} | {111} | {112} |
|---|---|---|---|---|
| $E_a$ (eV) | 0.90 | 0.92 | 1.26 | 1.06 |

While in general the simulation data for all the growth orientations support an exponential dependency on temperature, the values of $E_a$ exhibit noticeable differences. Substrates with a larger $E_a$ (e.g., for the {111} plane) are expected to have a growth pathway containing a higher barrier (that may indicate a nucleation process), and these substrates also typically suffer from a slower growth kinetics. Microscopically, this may be related to both the impingement rate and the surface stability of the atoms, which will be discussed with more details in Section II.B.

## 2. Concentration-dependency analysis

Since the temperature-dependent term in Eq. 1 can be evaluated with knowing the activation energy $E_a$, it is possible to numerically express $f(\Delta\text{Ge})$ by plugging the simulation data (i.e., the growth rate $r$ at each $\Delta\text{Ge}$ and $T$). These data points are collected in Fig. 3, with the four subfigures corresponding to the four substrate orientations.

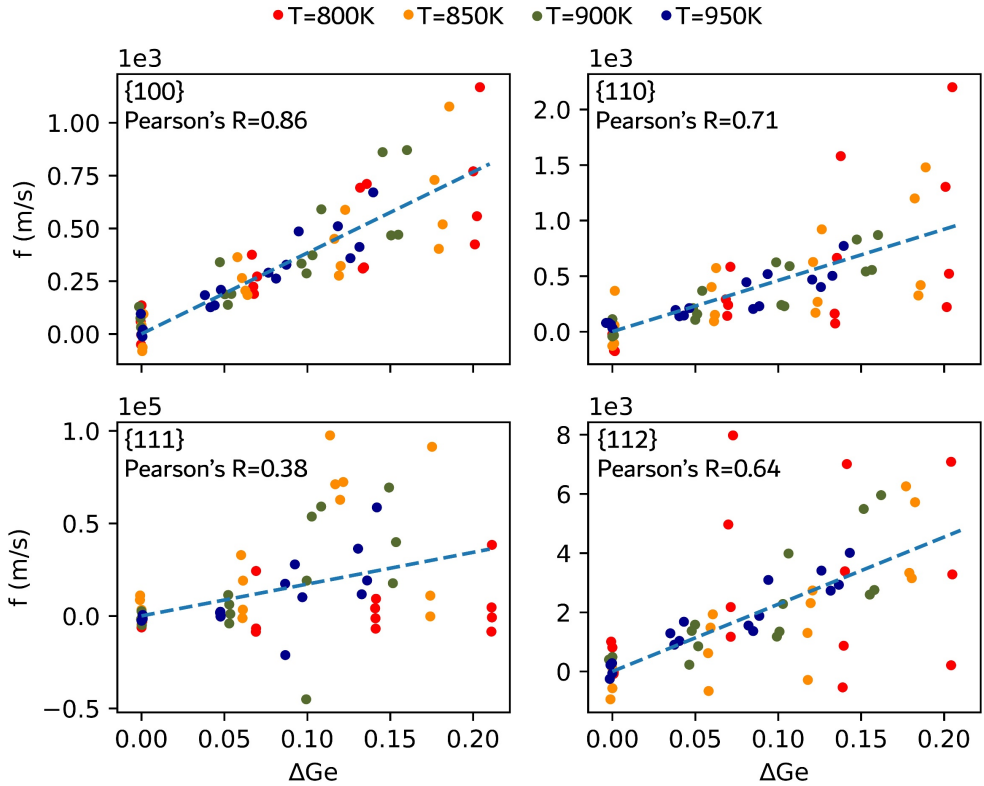

**Figure 3. The scatter plots of the calculated concentration-dependent factors $f(\Delta\text{Ge})$ and the proposed linear fitting functions.**

For each orientation, we calculated the value of Pearson's R to quantify the correlation between the concentration-dependent factor $f$ and the Ge supersaturation $\Delta\text{Ge}$. We found that $f$ of plane {100}, {110} and {112} are relatively more linearly correlated to $\Delta\text{Ge}$, with the Pearson's R equal to 0.86, 0.71, and 0.64 respectively. This suggests that the overall growth mode of these three planes may be expressed by Eq. 2,

$$r = A \cdot \Delta \text{Ge} \cdot \exp(-\frac{E_a}{k_B T}) \qquad \text{(Equation 2)}$$

where $A$ is a substrate orientation dependent constant: $A = 3,833$ m/s for {100}; $A = 4,605$ m/s for {110}; and $A = 22,701$ m/s for {112}, respectively. In contrast, the data of {111} plane exhibit a non-linear behavior with a much larger variation, suggesting that the growing mode of Ge crystal on {111} plane would be different from that of the other substrate orientations simulated in this study. In other words, for {111}, the temperature factor and the concentration factor may not be simply decoupled as described by Eq. 2. Hence, a further investigation into the microscopic growth mechanism on four substrates was conducted, aiming to describe the atom exchange phenomena near the solid-liquid interface (in Section II.B) and track the evolution of interface morphology (in Section II.C).

**B. Interface Atom Exchange**

A crystallization process accompanies with atom exchange events at the liquid-solid interface. To keep track of these events, the trajectories of individual Ge atoms in our simulations were recorded, where two types of Ge atoms could be categorized. The type 1 atoms stayed in a state of either liquid or solid during the whole simulation time; while the type 2 atoms experienced one or more times of state change, either from liquid to solid or from solid to liquid, which is clearly of more interest. For the type 2 atoms, three different behaviors could be further identified from their trajectories. (1) we found that some atoms were diffused from solid to liquid and stayed in the liquid phase till the end of the simulation. (2) some atoms solidified on the substrate at some time and became stable as part of the solid substrate, finally contributing to the Ge crystal growth. (3) we also observed that some other atoms crystallized first and stayed in a solid environment for a period, but finally diffused back into the liquid, whose process was referred as an "unstable crystallization". The behaviors of those atoms were presented as a set of histograms in Fig. 4&5, where hidden behind statistical patterns can be discovered for understanding the atom exchange process at the growth interface.

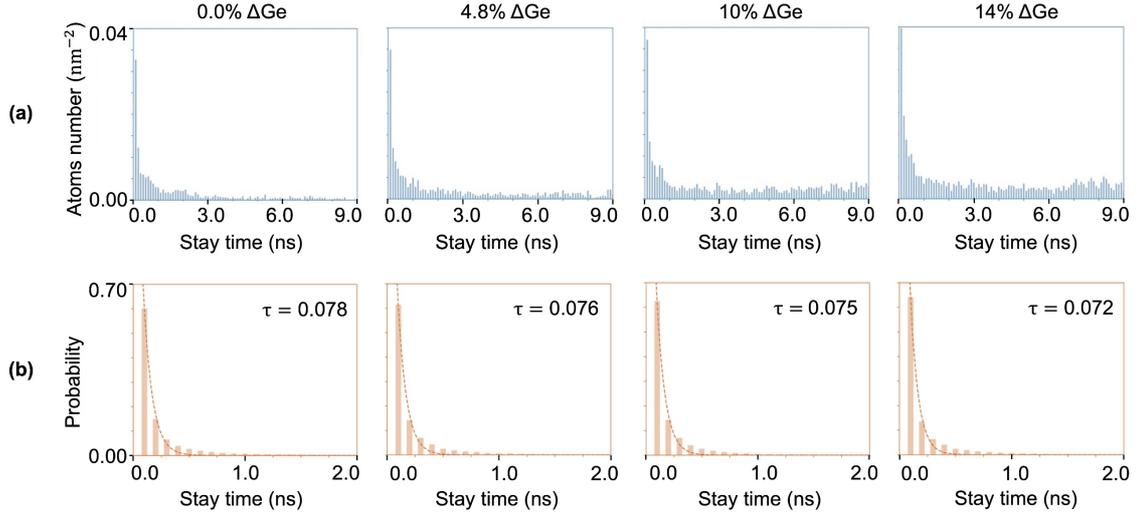

**Figure 4.** (a) The distribution of type 2-(2) atoms at the end of a 10-ns simulation with respect to their total time stayed on the {100} substrate. (b) The probability distribution of all the unsuccessful growth attempt $p$ with respect to the time that the type 2-(3) atoms stayed on the solid substrate $t$. $\tau$ is the parameter fitted with the exponential decay equation $p = A \cdot \exp(-\frac{t}{\tau})$.

Fig. 4(a) provides an overview of the Ge solidification process at 950K on the solid substrate of {100}. The four panels corresponded to the simulation data from four initial Ge supersaturation conditions. All the Ge atoms that crystallized from liquid at some time and stayed in solid till the end of the simulation were included in these diagrams, where the *x*-axis represents the time of a Ge atom staying in solid once crystallized, and the *y*-axis collects the counts normalized by the area of the liquid-solid interface. Clearly, the bars at larger *x* refer to the atoms that crystallized earlier with a longer stay time (in solid), and they are likely to have a higher probability to continuously be stable if the simulation time could be extended. The four diagrams in Fig. 4(a) demonstrate a similar distribution, i.e., a rapid decay in a short period followed by a long flat tail in a longer period. The atoms distributed in the flat tail were probably those atoms that were stably grown onto the substrates. Thus, longer and higher tails were seen for simulations at higher Ge supersaturations, consistent with the positive correlations found previously between the growth rate *r* and the Ge supersaturation ΔGe.

The initial rapid decay in Fig. 4(a) could be attributed to those "unstable crystallization" atoms that finally diffused back to liquid. Though the "unstable crystallization" did not contribute to the macroscopic crystal growth, it involved multiple solidification and liquidation processes of atoms at microscopic scale, where several important surface properties, such as the atom activity and stability, could be reflected. In Fig. 4(b), distributions of the atoms undergoing an "unstable crystallization" process were plotted, with the $x$-axis representing the atoms' stay time at the solid substrate and the $y$-axis showing the corresponding probability. The data were fitted with an exponential decay equation, where a constant $\tau$ can be obtained as a descriptor of the surface atom stability. Since $\tau$ slightly decreased with the increase of $\Delta$Ge, this suggests that the "unstably crystallized" atoms tend to leave the surface sooner under a larger growth driving force. Also, we found that all these atoms typically would not stay for over 2 ns. Then, 2 ns may serve as an empirical threshold value for differentiate "stable" and "unstable crystallization", such that if a Ge atom can maintain its crystalline state for the first 2 ns, it would be likely to be stable in the solid state.

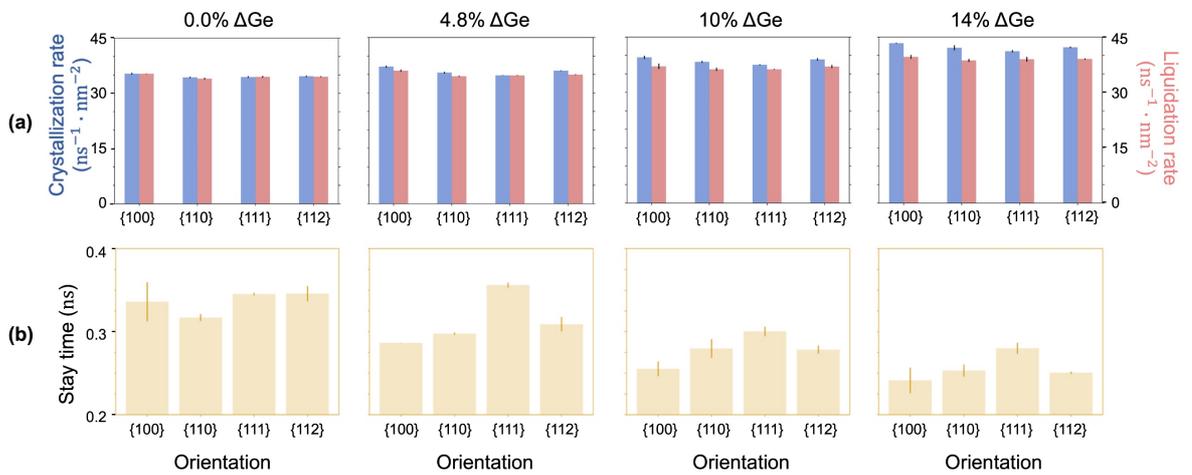

**Figure 5. (a) Histograms of crystallization rate versus liquidation rate for all type 2 atoms on different substrates under four Ge supersaturations. (b) Average stay time of unsuccessful crystallization attempts for type 2-(3) atoms on different substrates under four Ge supersaturations.**

Fig. 5(a) and 5(b) provide another angle to view this "unstable crystallization".

Fig. 5(a) gives information of how frequently the "unstable crystallization" happened per unit area on the four types of substrates for a set of Ge supersaturations. The blue bar counts the number of the transition from liquid to solid $\Gamma_{l \to s}$, and the pink bar counts the number of the transition from solid to liquid $\Gamma_{s \to l}$, so the difference between the two ($\Gamma_{l \to s} - \Gamma_{s \to l}$) refers to the number of successful attempts of crystallization. Then, $\Gamma_{l \to s}$ may be useful for describing the surface atom activity, and $\Gamma_{l \to s} - \Gamma_{s \to l}$ may serve as an indicator for surface atom stability. From Fig. 5(a), a few more observations could be achieved. First, the ratio of $\Gamma_{l \to s} - \Gamma_{s \to l}$ to $\Gamma_{l \to s}$ may reveal the crystallization efficiency at the microscopic level. As the absolute value of red and blue bar are all around 30 to 45 $\text{ns}^{-1} \cdot \text{nm}^{-2}$, the crystallization efficiency $\xi$ (defined as $\frac{\Gamma_{l \to s} - \Gamma_{s \to l}}{\Gamma_{l \to s}}$) would be low (see Tab. S1 for more details). Second, both $\Gamma_{l \to s}$ and $\Gamma_{s \to l}$ were enhanced with the increase of ΔGe, suggesting a positive correlation between ΔGe and atom activity. Third, the number of $\Gamma_{l \to s}$ for the {111} plane was typically the lowest when compared among different substrates, which may be further correlated to the surface morphology (see Section C for more details).

In Fig. 5(b), the averaged "unstable crystallization" time was calculated and compared among the cases of different substrate orientations and ΔGe. Generally, a larger supersaturation led to a less average time of "unstable crystallization", indicating a higher surface atom activity. Besides, purely focusing on the substrate effects, a longer stay time was typically seen for {111} plane. That is to say, during an "unstable crystallization", atoms on {111} plane are more difficult to leave the surface once attached, which may indicate a higher activation energy of diffusion, and again could be related to the special {111} surface morphology.

### C. Liquid-Solid Interface Morphology

From above, it can be seen that the surface morphology of the Ge solid interface is important for its role in influencing the probability of atom exchange at the liquid-solid interface. One possible parameter to describe the surface morphology could be its roughness. The surface roughness factor $\eta$ was calculated by summing up the square of the in-plane ($x$-$y$ plane) gradient evaluated at the grid points of a surface mesh, which was constructed from a selection of Ge solid atoms at the top surface. As shown in Eq. 3, $z_{i,j}$ is the height of the point ($i$, $j$) on the surface mesh.

$$\eta = \sum_i \sum_j \left(\frac{\partial z_{i,j}}{\partial x_{i,j}}\right)^2 + \left(\frac{\partial z_{i,j}}{\partial y_{i,j}}\right)^2 \quad \text{(Equation 3)}$$

When ΔGe = 0, as shown in Fig. 6(a), the {111} plane can be considered as a smooth surface at 950K, since $\eta$ was close to 0 during the entire simulation. While the {111} plane was kept a close-to-perfect close-packed surface, the other three planes developed a similar degree of roughness at equilibrium. But as the liquid became saturated (ΔGe% = 10 at 950K), as shown in Fig. 6(b), the surface roughness of the {111} plane started to vary with time. At the meantime, the surface roughness of {100}, {110} and {112} did not change much with the increase of ΔGe, presumably due to an intrinsic rough nature of these interface orientations. In addition, it should be noted that without the growth driving force, the roughness of a given plane could also be affected by temperature (see Fig. S2 for more details).

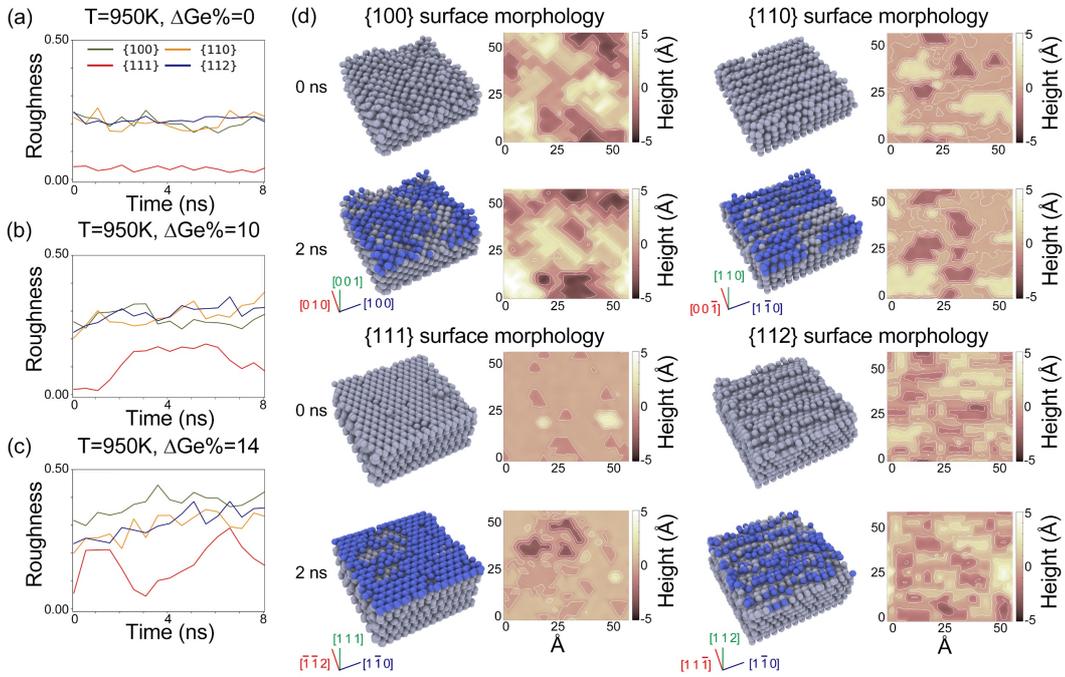

**Figure 6. Surface roughness of four types of liquid-solid interfaces (a) in equilibrium ($\Delta \text{Ge}\% = 0.0$), (b) at $\Delta \text{Ge}\% = 10$, and (c) at $\Delta \text{Ge}\% = 14$, with keeping the temperature at 950K. (d) 3D views and contour maps showing the four**

**interfaces at 950K and ΔGe% = 14. Atoms in grey were Ge atoms originally on solid substrates and atoms in purple-blue were the newly grown atoms during a 2ns simulation. The average height of surface atoms of each plane was set to 0.**

In Fig. 6(d), the contour maps of surface height were provided for comparing the surface morphologies and growth modes among the four substrate orientations at 950K and ΔGe% = 14. For {100} plane, several layers of atoms were initially exposed, and the new Ge atoms can grow on both the upper layers of atoms and the lower layers, leading to a marginally increasing $\eta$ during the simulation (Fig. 6(c)). For {110} plane, the initial surface morphology was approximately composed of 3 incomplete layers of atoms. Then during a growth of 2 ns, the new Ge atoms attached to the surface and formed several large patches. For {112} plane, several ridges and steps were seen to align along the ⟨110⟩ direction, which may provide preferrable sites for the new Ge atoms to join. For {111} plane, while initially most of the atoms formed a close-packed plane with only a few atoms stacked on the 2$^{nd}$ layer, a new close-packed layer was formed during a MD simulation of 2 ns. Interestingly, according to Fig. 6(c), a two-step growth along the ⟨111⟩ orientation was indicated: the atoms firstly formed a number of Ge seeds, like small islands scattered on the initially flat surface, which increased the surface roughness. Then, new atoms started attaching onto these islands. Finally, the islands connected into a continent and completed a new close-packed crystal layer. According to Fig. 6(c), this layer-by-layer crystallization process led to a periodic change of $\eta$, as we found the surface roughness of the {111} plane rose first and dropped close to 0 (indicating the completeness of the layer) at around 3.0 ns. Later, when another new layer of atoms was initiated on {111} plane, presumably in the form of several islands, the surface roughness started to increase, followed by the second drop of $\eta$ at around 6.5 ns. This helps explain the non-uniform growth rate of {111} system at 950K shown in Fig. 1, where the slow-growth periods may present the incubation and nucleation stage, and the rapid-growth periods may reflect the filling of atoms on an "activated" {111} surface.

## III. Discussion

In this research, we systematically investigated the mechanisms of Au-catalyzed Ge crystal growth by molecular dynamics simulations. From the temperature-dependency analysis, we estimated the overall activation energies of Ge crystal on the four common crystallographic orientations of Ge substrates. Our concentration-dependency analysis demonstrated a positive linear correlation between the Ge supersaturation and the VLS crystal growth rates on {100}, {110} and {112} planes, eventually allowing for prediction on the growth rate as a function of temperature and supersaturation for these substrates, while a more complex behavior was identified for the {111} growth. Microscopically, the behaviors of surface Ge atoms were understood by the classifications of "stable" and "unstable" crystallizations, while an empirical cut-off value of staying time may help quantify the surface stability of the atoms. It was also found that for the atoms undergoing "unstable crystallization", an increased supersaturation led to a decrease of the average stay time, maybe due to an increased surface activity. Finally, a multi-step mechanism of {111} growth was revealed by analyzing the evolution of a surface roughness parameter $\eta$ during the VLS growth. Though the surface roughness is intrinsically determined by the stacking pattern of the atoms, it is observed to vary with temperature and Ge supersaturation, exerting influence on the instantaneous crystal growth behaviors of Ge.

It should be noted that several improvements can be made to produce more accurate simulation data and predictions. Firstly, as we did not allow a dynamical adjustment of the number of Ge atoms during the simulation, the Ge concentration may be dropped by up to 2% with the consumption of Ge atoms in the liquid. This may possibly cause errors in our predictions. For future study, a more sophisticated algorithm may be developed to mimic the atom exchanges at the liquid/vapor interface, without inducing noticeable artificial perturbations to the system. Secondly, for a more accurate prediction on the instantaneous growth rate, especially for {111} planes, model parameters such as the activation energy may be set to dependent on surface roughness, to account for the surface morphology evolution during the layer-by-layer growth. Overall, this research revealed a detailed microscopic picture of Au-catalyzed Ge VLS growth under a wide range of temperature and supersaturation conditions, with systematic analyses of the orientation dependent growth behaviors. These results could inspire the mechanistic study of catalyzed crystal growth of similar systems and provide useful guidelines for controllable nano-synthesis of high-quality semiconductor materials.

## IV. Methods

### A. MD simulation of Ge Crystal Growth

All the MD simulations were carried out by the Large-scale Atomic/Molecular Massively Parallel Simulator (LAMMPS)[28] using the MEAM potential[26] for the Au-Ge binary system. To ensure the comparability among configurations with different crystallographic orientations of Ge substrates, we controlled the size of the simulation cell to be similar with each other. The total number of atoms in one cell was set to around 12,000, with a width and length of around 6nm. The solid substrate contained more than 3 layers of unit cells, with the bottom layer fixed in order to prevent potential drift of the box. The simulations were run under the periodic boundary condition along horizontal directions, at a time step of 0.5 ns with an overall time of 10 ns. A NVT ensemble implemented as a Nose-Hoover thermostat was applied to control the temperature of the system. An embedded order parameter $q_3$ proposed in Steinhardt's paper[29] was adopted to determine the state of an atom (liquid or solid) in our simulations.

We set $q_3$=0.50 as the threshold for all the simulations, such that the Ge atoms that have an order parameter $q_3 < 0.50$ were categorized as liquid, while those whose $q_3 \geq 0.50$ were grouped into solid.

**B. Identification of the liquid-solid interface**

Based on the $q_3$ value, all the solid Ge atoms can be identified at every timestep. This allows to keep track of the surface morphology of the growth interface by the construction of a surface mesh of Ge atoms in OVITO[30], using the alpha-shape method of Edelsbrunner and Mücke[31]. In our simulations, the Ge atoms close to the upper generated surface mesh were selected as the surface atoms, for the calculation of surface roughness using Eq. 3.

**Acknowledgments**

YW would like to acknowledge funding from the National Natural Science Foundation of China (Grant No. 5210020146). ZW would like to acknowledge the support from the Shanghai Jiao Tong University Participation in Research Program (PRP).

# Supplementary Information

# Revealing atomistic mechanisms of gold-catalyzed germanium growth using molecular dynamics simulations

Zhiyi Wang[1], Jixin Wu[2], Yanming Wang[1, *]

[1] *University of Michigan - Shanghai Jiao Tong University Joint Institute, Shanghai Jiao Tong University, 800 Dongchuan Rd., Minhang District, Shanghai, 200240, P. R. China*
[2] *School of Materials Science and Engineering, Huazhong University of Science and Technology, 1037 Luoyu Rd., Hongshan District, Wuhan, Hubei, 430074, P. R. China.*

[*]Corresponding author, yanming.wang@sjtu.edu.cn (Y. Wang)

**Calculation of Ge crystal growth rate**

The original data of newly crystalline atoms on the solid substrates were collected as scattered values with respect to the simulation timestep (in ps). To convert the number of crystalline atoms into the total length of crystal growth, we have to calculate how each atom contribute to the growth on average. For the diamond structure of Ge, the volume of a common unit cell is shared by 8 atoms and then by dividing the volume with the cross-section area of the Ge substrate, we could get the growth length contribution of each Ge atoms. The scattered points were plotted in Fig.S1 and through interpolation, we could smoothen the growth process. Then the growth rate could be found by calculating the slope of the tangent line of the curves. For each growth curve, four growth rates at 2000ps, 4000ps, 6000ps and 8000ps were recorded with their corresponding instantaneous $\Delta Ge\%$.

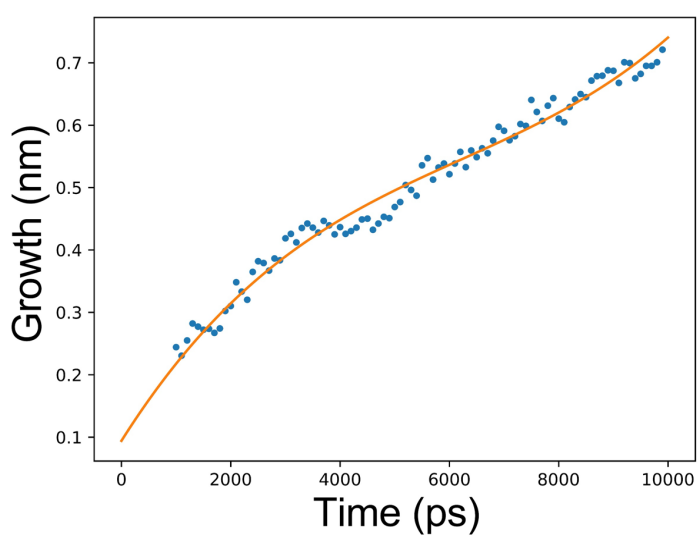

Figure S1. The original scattered growth data on crystallographic orientation {100} with initial $\Delta Ge\%=9.7\%$ at 950K and the smooth growth curve after the interpolation.

**Estimate of crystallization efficiency**

The microscopic crystallization efficiency $\xi$ could be evaluated by Eq. S1:

$$\xi = \frac{\Gamma_{l\to s} - \Gamma_{s\to l}}{\Gamma_{l\to s}} \qquad \text{(Equation S1)}$$

All the $\xi$ values at 950K are shown in Table S1.

Table S1. $\xi$ values at various substrate orientation and Ge supersaturation conditions.

|   | 0.0% | 4.8% | 10% | 14% |
|---|---|---|---|---|
| {100} | 0 | 0.0292 | 0.0618 | 0.0851 |
| {110} | 0 | 0.0280 | 0.0519 | 0.0821 |
| {111} | 0 | 0.0003 | 0.0325 | 0.0529 |
| {112} | 0 | 0.0300 | 0.0494 | 0.0748 |

**Temperature dependency of surface roughness**

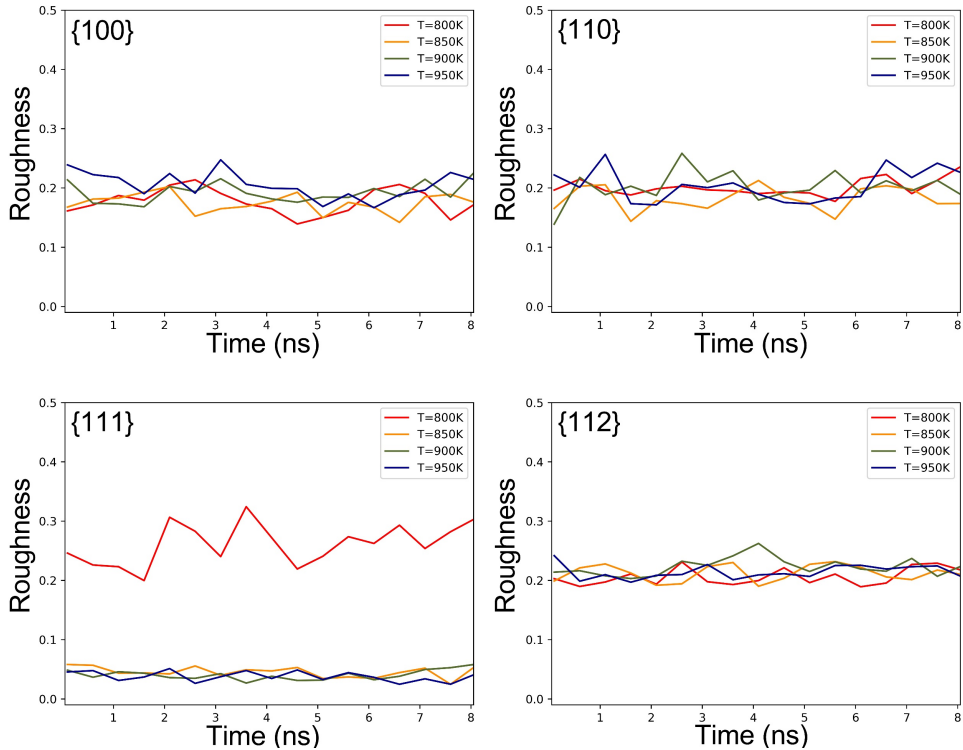

Figure S2. Surface roughness of different crystallographic orientations at ΔGe%=0 with respect to four different temperature.

Based on the curves in Fig. S2, the surface roughness of {100}, {110}, {112} did not show a clear trend with the increase of temperature. However, at 800K, {111} showed a rougher surface comparing to the surface under 850K, 900K, 950K. The rise of the roughness came from those crystallite atoms pinned onto the {111} flat surface, but in reverse, the thermodynamic fluctuation at 800K makes atoms difficult to overcome the energy barrier to liquidized, which could finally lead to a rougher {111} surface at 800K.